\documentclass[
    ,final            
  ]
  {aipproc}

\layoutstyle{6x9}

\begin{document}

\title{Overview of Charm Physics at RHIC}


\author{M.J. Leitch}{
  address={Los Alamos National Laboratory, Los Alamos NM 87545 USA, leitch@lanl.gov}
}



\begin{abstract}
Heavy-quark production provides a sensitive probe of the gluon structure of
nucleons and its modication in nuclei. It is also a key probe of the
hot-dense matter created in heavy-ion collisions. We will discuss the
physics issues involved, as seen in quarkonia and open heavy-quark
production, starting with those observed in proton-proton collisions. Then
cold nuclear matter effects on heavy-quark production including shadowing,
gluon saturation, energy loss and absorption will be reviewed in the
context of recent proton-nucleus and deuteron-nucleus measurements. Next
we survey the most recent measurements of open-charm and $J/\psi$s in
heavy-ion collisions at RHIC and their interpretation. We discuss the
high-$p_T$ suppression and flow of open charm in terms of energy loss and
thermalization and, for $J/\psi$, contrast explanations in terms of screening
in a deconfined medium vs. recombination models.
\end{abstract}

\maketitle


\section{Charm Production in p+p Collisions at RHIC}

Gluon fusion dominates the production of quarkonia, but the configuration of the
produced state and how it hadronizes remain uncertain. Absolute cross sections
can be reproduced by NRQCD models that involve a color octet state\cite{beneke},
but these models predict transverse polarization of the $J/\psi$ at large $p_T$ that is
not seen in the data\cite{beneke2}. A general complication in understanding $J/\psi$ results is
the fact that $\sim$40\% of the $J/\psi$s come from decays of higher mass resonances
($\psi'$ and $\chi_C$)\cite{abt} - a feature that may contribute to the lack of polarization seen.
One exception to this feature is the maximal transverse polarization observed
for the $\Upsilon_{2S+3S}$ states\cite{cnbrown}; where the lack of feed-down
for these states may allow the polarization to persist.

$J/\psi$ cross section measurements for p+p collisions at  $\sqrt{s}=200$ GeV from
PHENIX\cite{phenix_jpsi_pp} are shown in Fig.~\ref{fig:sigpp}.
These results, based on approximately 500 $J/\psi$s from the 2003 run, provide the baseline
for both CNM studies in d+Au collisions and QGP studies in A+A collisions at RHIC,
and are presently one of the limiting factors in obtaining precise nuclear modifications.
However p+p data from the 2005 and 2006 runs will soon improve this baseline significantly
with over 40,000 $J/\psi$s.

Open charm measurements at RHIC suffer from large systematics and statistical uncertainties
due to the statistical subtraction methods that are used. Measurements by PHENIX and
by STAR differ substantially\cite{phenix_akiba_charm} on the size of the charm cross section.
Also the measured cross sections lie substantially higher than current theoretical predictions
as shown in Fig.~\ref{fig:charm_pp_nrqcd}.

\begin{ltxfigure}[tbh]
\begin{minipage}[b]{0.44\linewidth}
  \centering
  \includegraphics[width=0.65\textwidth]{figures/sigpp.eps}
  \caption{$J/\psi$ cross section vs rapidity for 200 GeV p+p collisions at
  RHIC\cite{phenix_jpsi_pp}.}
  \label{fig:sigpp}
\end{minipage}
\hspace{0.2cm}
\hfill
\begin{minipage}[b]{0.38\linewidth}
  \centering
  \includegraphics[width=0.9\textwidth]{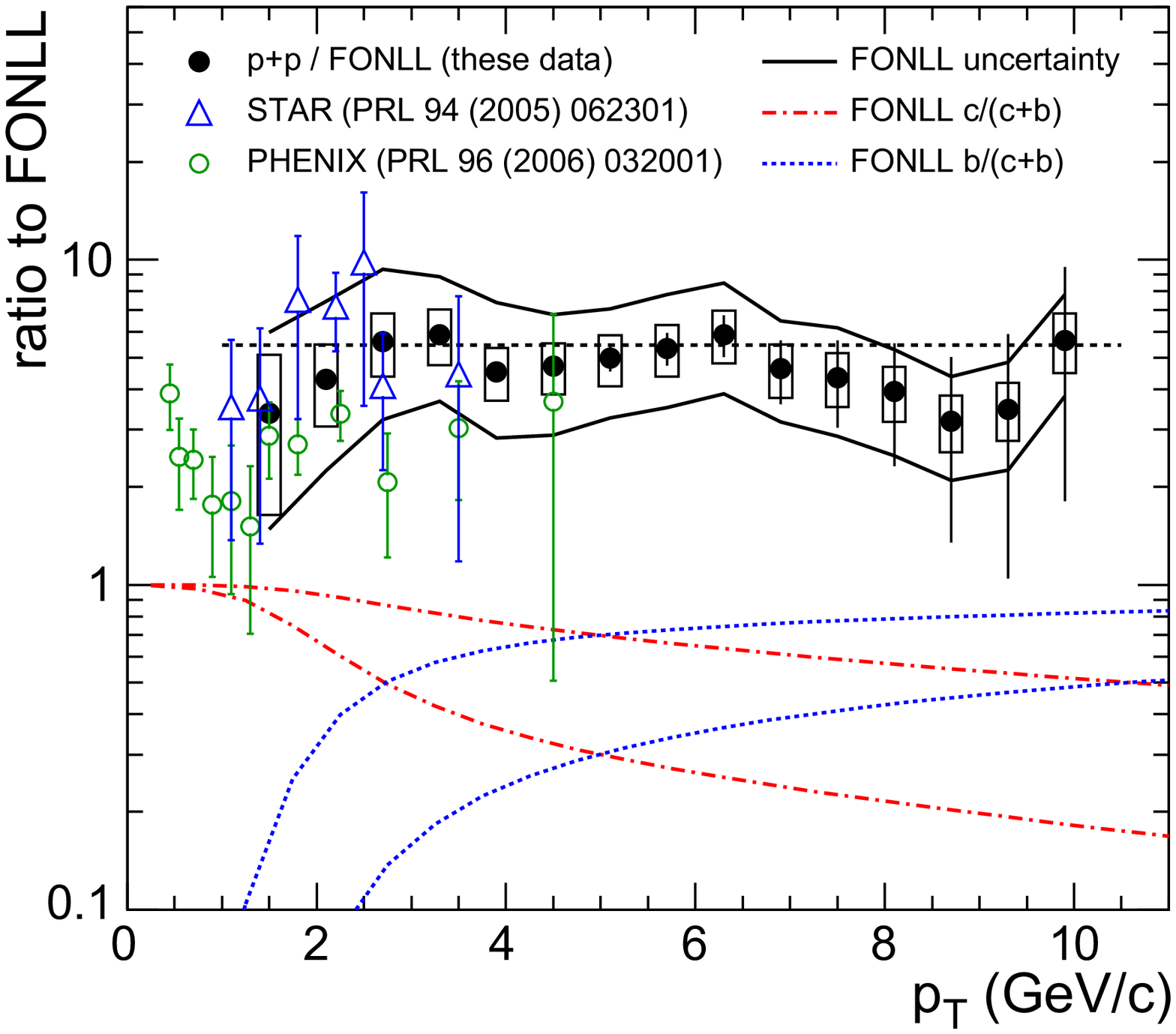}
  \caption{Open charm plus beauty cross section from prompt electrons divided by FONLL theory
  vs $p_T$ for 200 GeV p+p collisions at RHIC\cite{star_open_heavy}.}
  \label{fig:charm_pp_nrqcd}
\end{minipage}
\end{ltxfigure}

\section{Nuclear Effects on Charm}

When quarkonia are produced in nuclei their yields per nucleon-nucleon collision are known
to be significantly modified. This modification, shown vs. $x_F$ in
Fig.~\ref{fig:axfd_sardinia} for 800 GeV p+A
fixed target measurements and in Fig.~\ref{fig:rda} at RHIC energy,
is thought to be due to several CNM effects
including gluon shadowing, initial-state gluon energy loss and multiple scattering, and
absorption (or dissociation) of the $c\bar{c}$ in the final-state before it can form a $J/\psi$.

\begin{ltxfigure}[tbh]
\begin{minipage}[b]{0.38\linewidth}
  \centering
  \includegraphics[width=0.8\textwidth]{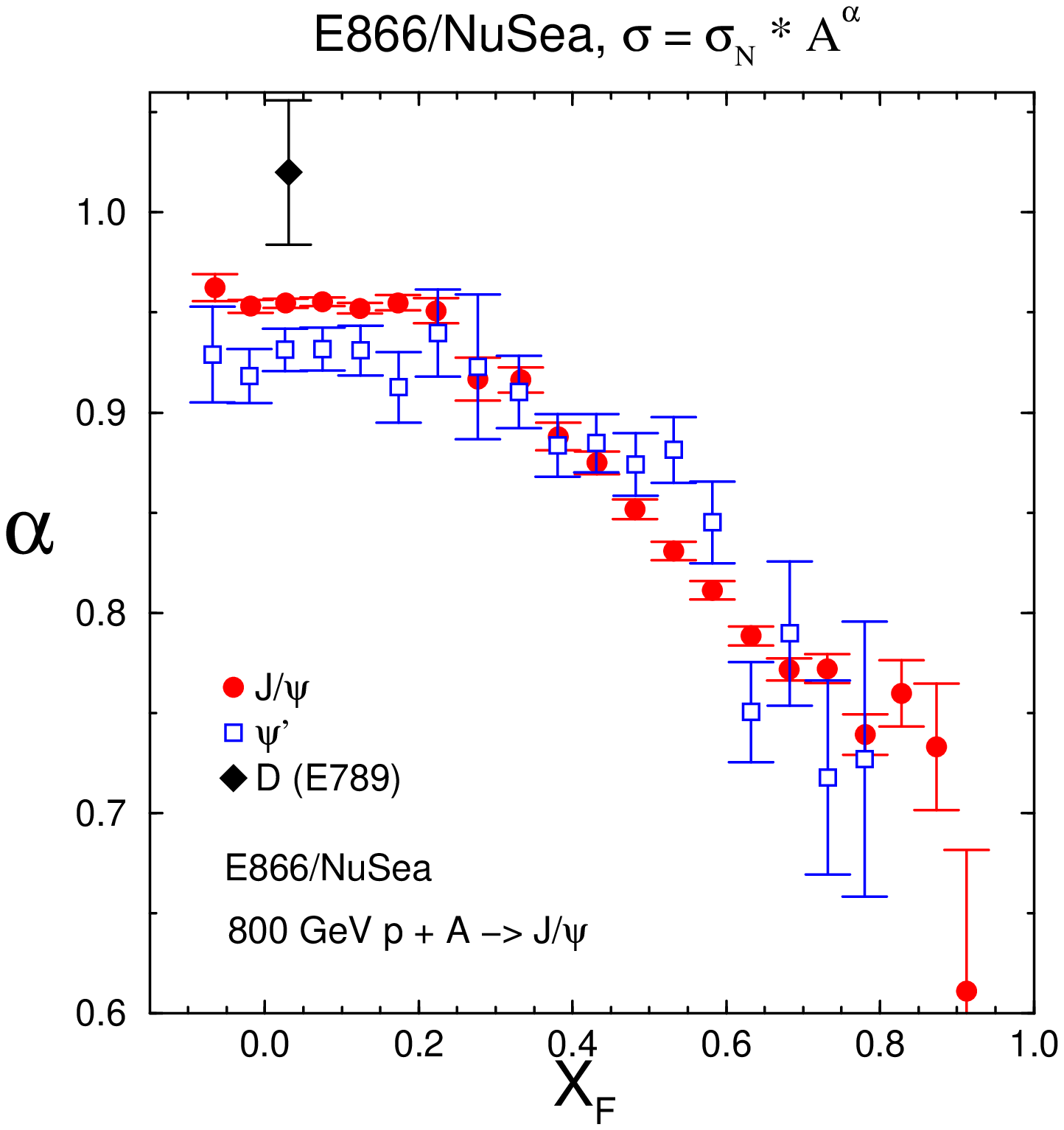}
  \caption{Nuclear moification factor $\alpha$ vs $x_F$ for $J/\psi$ and $\psi'$
  production in $\sqrt{s} = 38$ GeV collisions in E866/NuSea\cite{e866jpsi},
  and for $D^0$ from E789\cite{e789d}.}
  \label{fig:axfd_sardinia}
\end{minipage}
\hspace{0.2cm}
\hfill
\begin{minipage}[b]{0.40\linewidth}
  \includegraphics[width=0.75\textwidth]{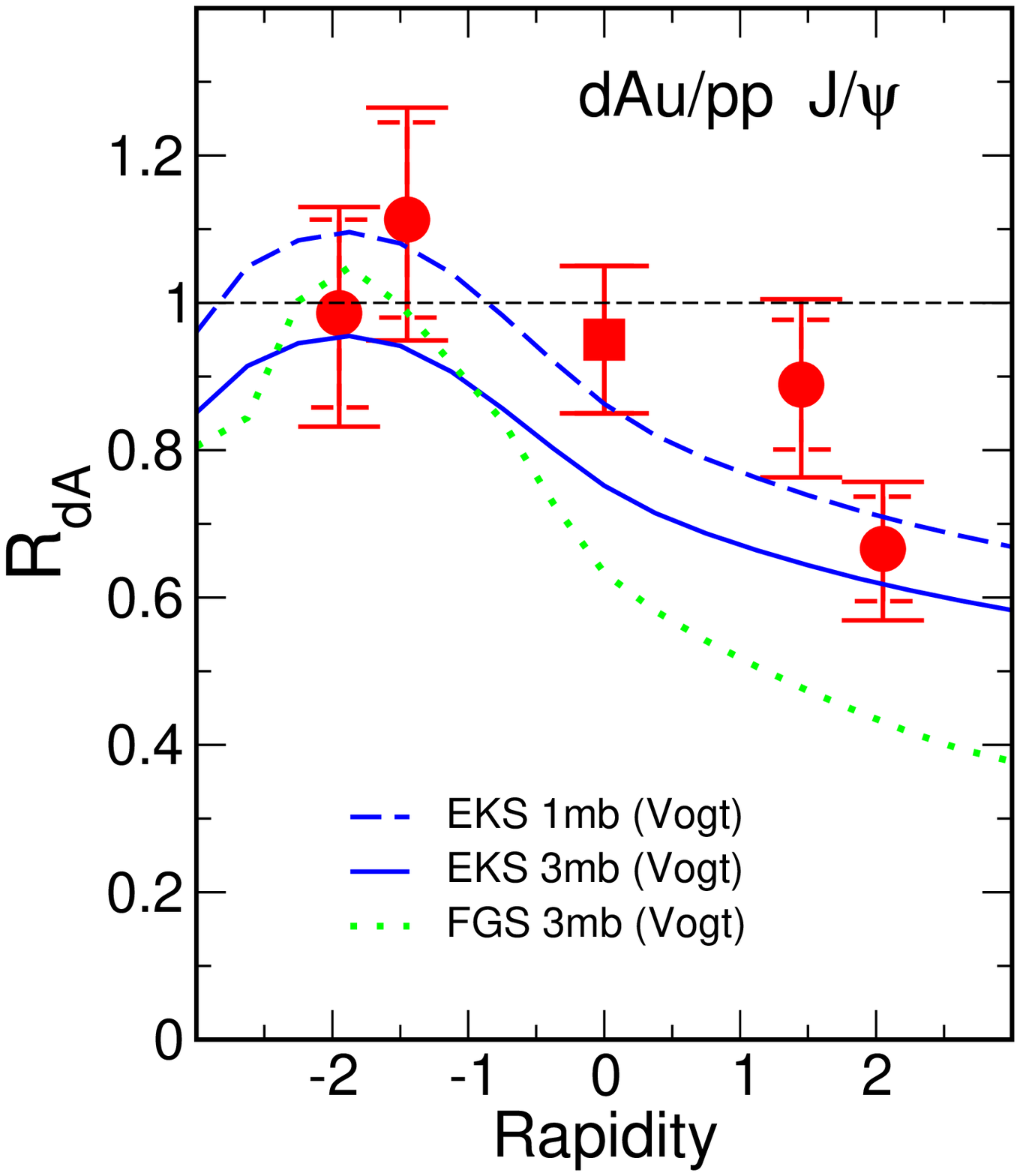}
  \caption{Rapidity dependence of the $J/\psi$ nuclear modification factor, $R_{dAu}$ for 200 GeV d+Au
  collisions at RHIC\cite{phenix_jpsi_pp}.}
  \label{fig:rda}
\end{minipage}
\end{ltxfigure}

Shadowing is the depletion of low-momentum partons (gluons in this case) in a nucleon
embedded in a nucleus compared to their population in a free nucleon. The strength of the
depletion differs between numerous models by up to a factor of three. Some models are based
on phenomenological fits to deep-inelastic scattering and Drell-Yan data\cite{eks}, while others
obtain shadowing from coherence effects in the nuclear medium\cite{fgs,boris_shadowing}.
In addition, models such as the Color Glass Condensate (CGC)\cite{cgc} yield shadowing through
gluon saturation pictures where the large gluon populations at very small x in a nucleus
generate a deficit of gluons at small x.

In the final state, the produced $c{\bar{c}}$ can be disassociated or absorbed
on either the nucleus itself, or on light co-moving partons produced when the
projectile proton or deuteron enters the nucleus. The latter is probably only important
in nucleus-nucleus collisions as the number of co-movers created in a p+A or d+A
collisions is small.

\begin{ltxfigure}[tbh]
\begin{minipage}[b]{0.56\linewidth}
  \centering
  \includegraphics[width=0.8\textwidth]{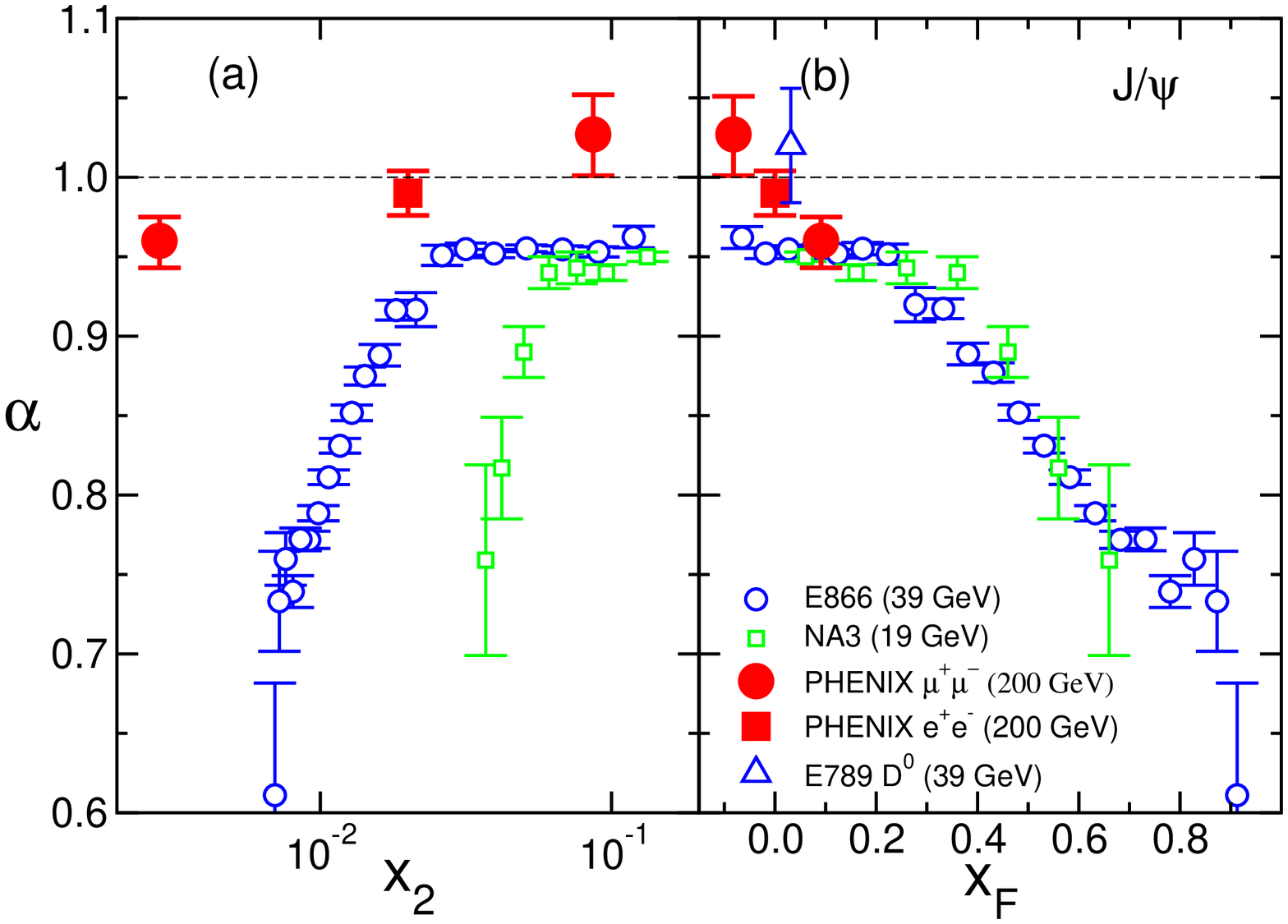}
  \caption{Test of scaling vs $x_2$ and $x_F$ for $J/\psi$ suppression data for
  three different collision energies. Data is from Refs.\cite{e866jpsi,na3,phenix_jpsi_pp}}
  \label{fig:alpha_x2xf_d}
\end{minipage}
\hspace{0.2cm}
\hfill
\begin{minipage}[b]{0.40\linewidth}
  \includegraphics[width=0.9\textwidth]{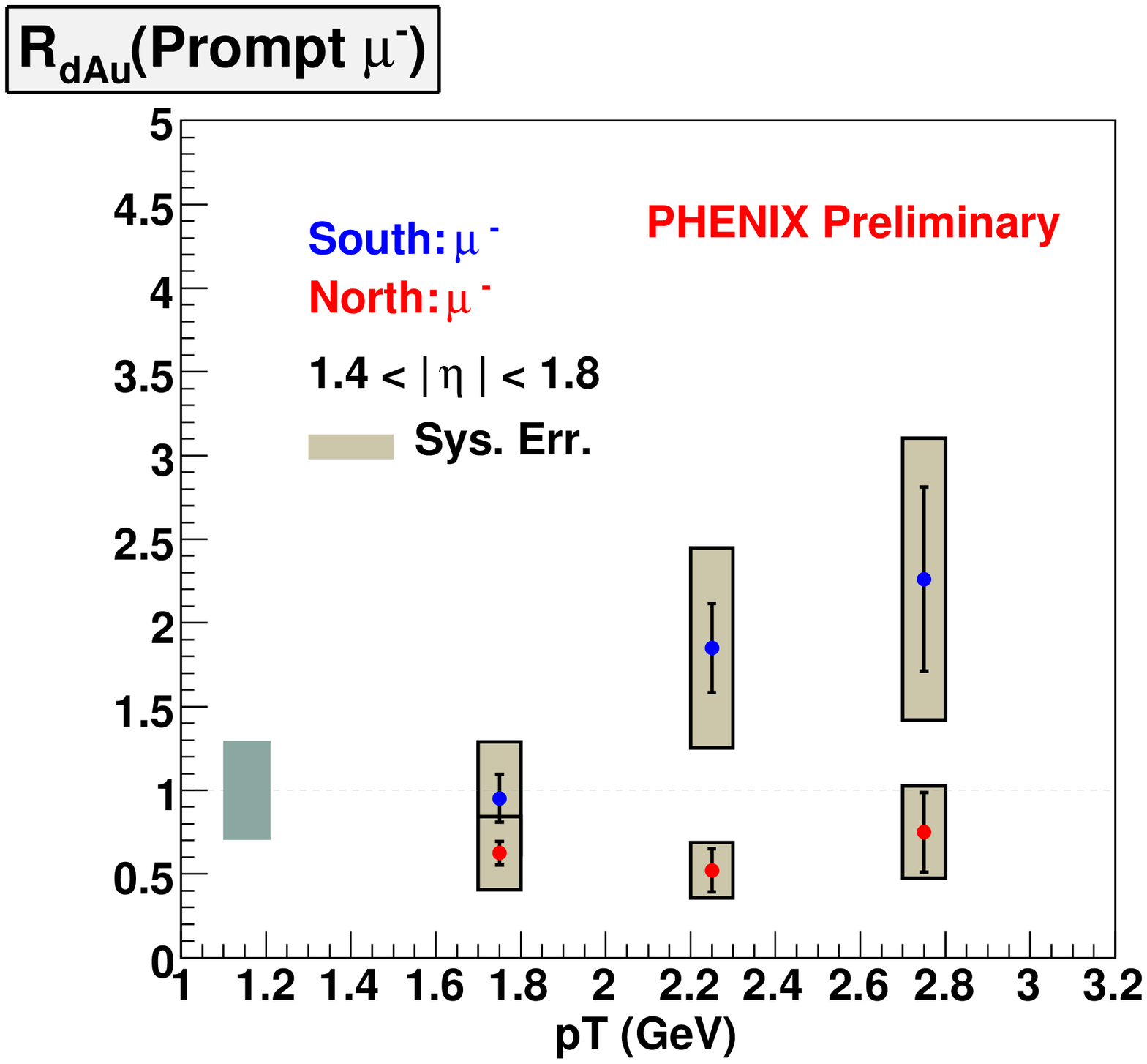}
  \caption{Nuclear dependence of heavy quark suppression vs $p_T$ from single muons in PHENIX.}
  \label{fig:charm_forward}
\end{minipage}
\hfill
\end{ltxfigure}

However, $J/\psi$ suppression in p(d)+A collisions remains a puzzle given that one does not
find a universal suppression vs $x_2$ as would be expected from shadowing, Fig.~\ref{fig:alpha_x2xf_d}a; 
while vs. $x_F$ the dependence is similar for all energies, Fig.~\ref{fig:alpha_x2xf_d}b. This apparent $x_F$
scaling supports explanations that involve initial-state energy loss or Sudakov
suppression\cite{sudakov}.

On the open-charm front, there are no substantial modifications seen at central rapidity in d+Au
collisions, but for forward rapidity (shadowing region) - as shown in Fig.~\ref{fig:charm_forward} - substantial
suppression is seen, while some enhancement is see at backward rapdity (Au-going direction).

\section{$J/\psi$ in Heavy-ion Collisions - A Quark Gluon Plasma Signature?}

One of the leading predictions for the hot-dense matter created in high-energy heavy-ion
collisions was that if a deconfined state of quarks and gluons is created, i.e. a
quark-gluon plasma (QGP), the heavy-quark bound states would be screened by the
deconfined colored medium and destroyed before they could be formed\cite{matsui_satz}.
This screening
would depend on the particular heavy-quark state, with the $\psi'$ and $\chi_C$ being dissolved
first; next the $J/\psi$ and then the $\Upsilon$'s only at the highest QGP temperatures. The CERN SPS
measurements\cite{na50} showed a suppression for the $J/\psi$ and $\psi'$ beyond what was expected from CNM
effects - as represented by a simple absorption model constrained to p+A data. In
addition to explanations involving creation of a QGP, a few theoretical models\cite{capella}  were also
able to explain the data without including a QGP, so the evidence that a QGP was formed
was controversial.

The first measurements from PHENIX at RHIC in 2004 are beginning to yield results - see
Fig.~\ref{fig:aa_cent_eks_band} for preliminary results for Au+Au and Cu+Cu collisions\cite{phenix_qm05_jpsi}.
First it is important
to understand what the normal CNM $J/\psi$ suppression should look like in these A+A
collisions. This is illustrated by the blue error bands for A+A collisions in
Fig.~\ref{fig:aa_cent_eks_band}
which represent identical theoretical calculations to the analogous blue error band in
Fig.~\ref{fig:rdau_eks98_band} for d+Au collisions. As can be seen the present d+Au data lacks enough precision
to provide a good constraint on the CNM effects. As a result
it is difficult to be very quantitative about the amount of "anomalous" suppression
observed in A+A collisions, although there does seem to be a clear suppression beyond CNM for the most
central collisions.

\begin{ltxfigure}[tbh]
\begin{minipage}[b]{0.35\linewidth}
  \centering
  \includegraphics[width=0.8\textwidth]{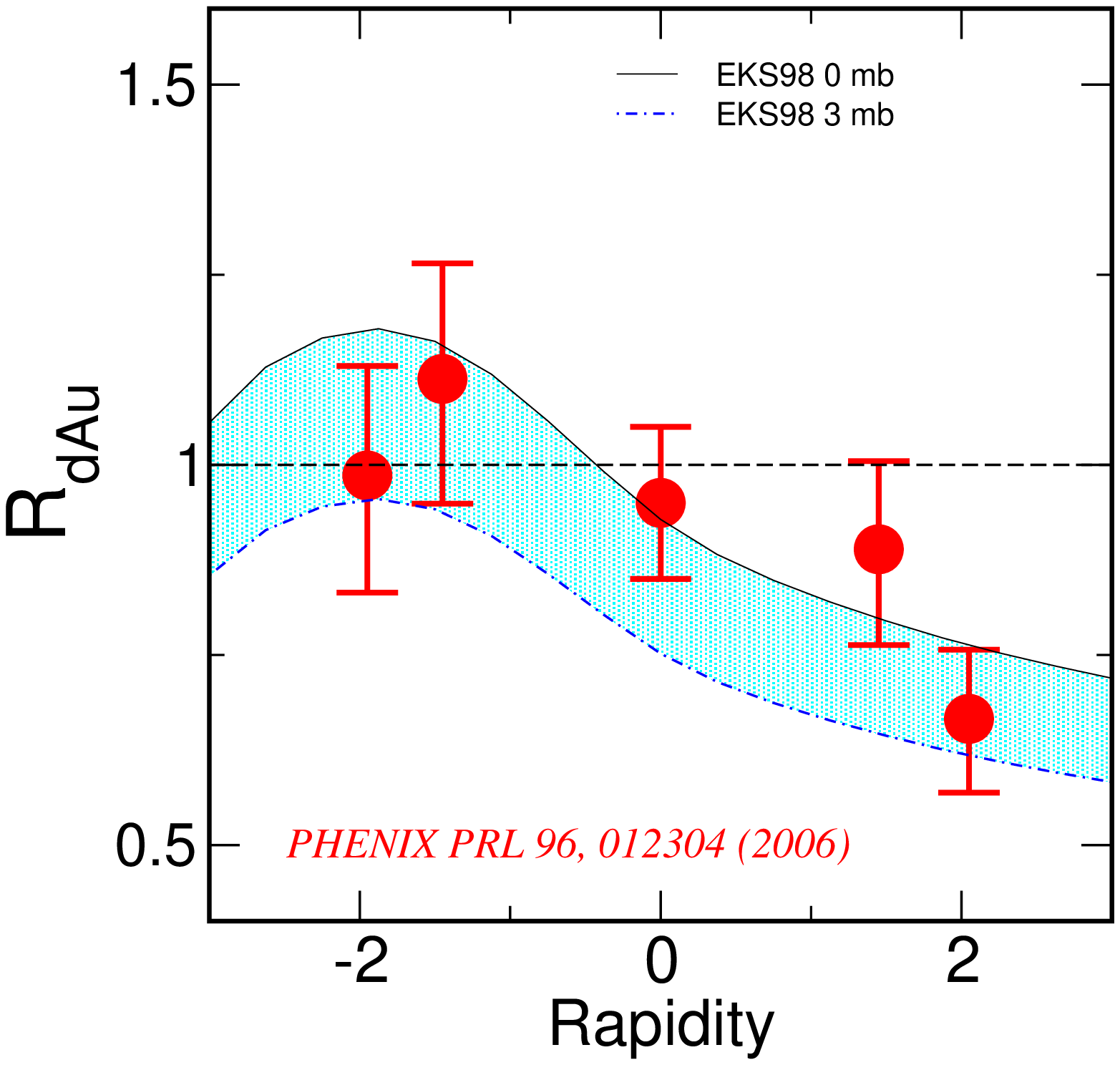}
  \caption{Results for $J/\psi$ suppression in d+Au collisions\cite{phenix_jpsi_pp} compared to a theoretical calculation
  that includes absorption and EKS shadowing\cite{vogt_cnm}.}
  \label{fig:rdau_eks98_band}
\end{minipage}
\hspace{0.2cm}
\hfill
\begin{minipage}[b]{0.52\linewidth}
  \includegraphics[width=0.8\textwidth]{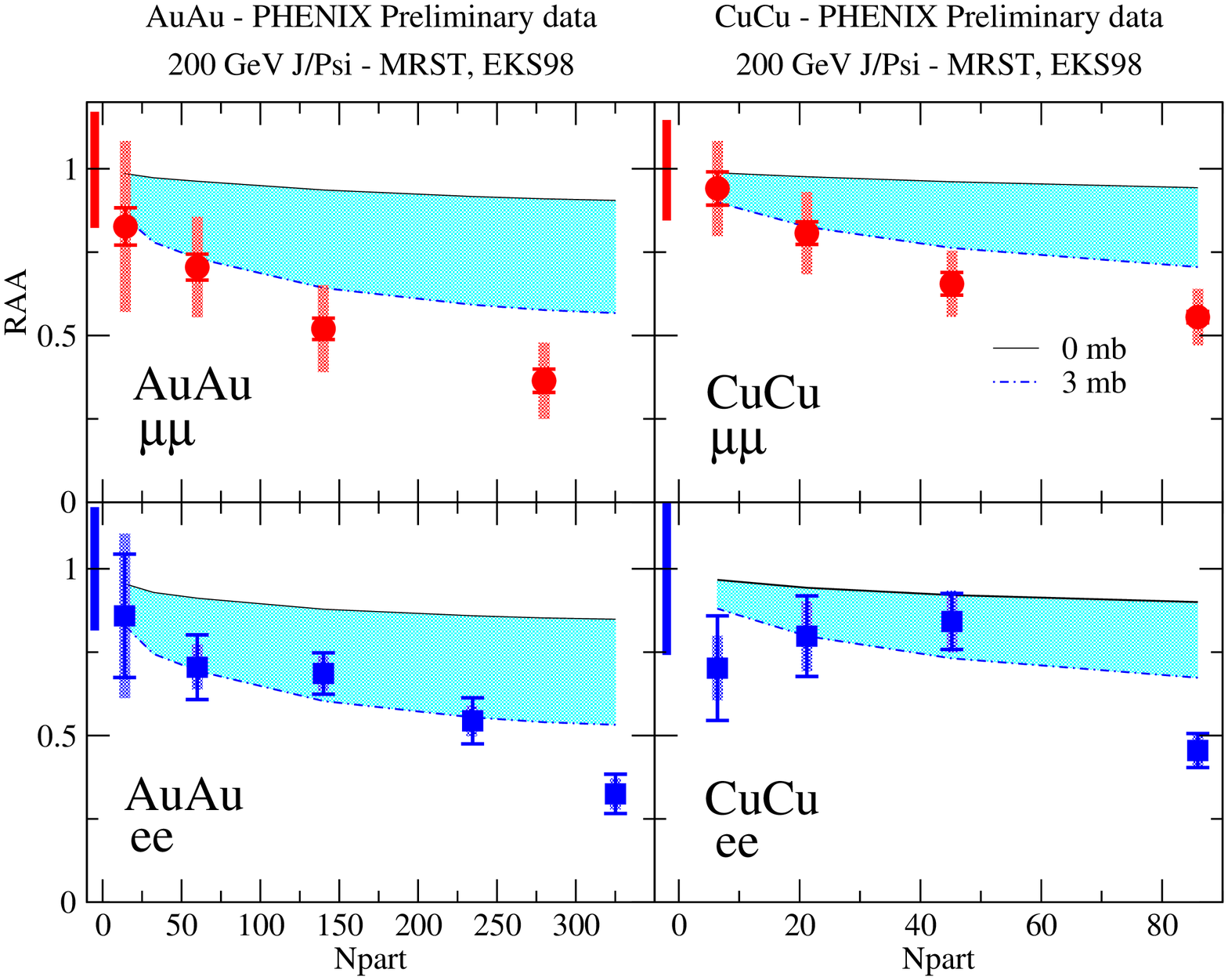}
  \caption{$J/\psi$ suppression in Au+Au and Cu+Cu collisions for forward rapdidity
  and central rapdity\cite{phenix_qm05_jpsi} compared to predictions for CNM from the same calculations
  as shown in Fig.~\ref{fig:rdau_eks98_band}\cite{vogt_cnm}.}
  \label{fig:aa_cent_eks_band}
\end{minipage}
\end{ltxfigure}

On the other hand, all of the models\cite{capella, satz,rapp}  that were successful in describing the lower
energy SPS data over-predict the suppression compared to the preliminary data at RHIC -
unless a "regeneration" mechanism is added as was done by Rapp\cite{rapp} and by Thews\cite{thews}.
The regeneration models assert that if the total production of charm is high enough then
densities in the final state will be sufficient to have substantial formation of $J/\psi$s
from the large number of independent charm quarks created in the collision. This production mechanism
was almost insignificant at SPS energies but at RHIC may be substantial. This leads to
a scenario in which strong screening or dissociation by a
very high-density gluon density occurs to a level of suppression stronger than the
RHIC data shows, but the regeneration mechanism compensates for this and brings the net
suppression back up to where the data lies. This is shown in Fig.~\ref{fig:raa_npart_supp_rapp}.

An alternative interpretation of the preliminary results, sequential screening, is given
by Karsch, Kharzeev and Satz\cite{karsch}. In this picture, they assume that the $J/\psi$ is never
screened, as supported by recent Lattice QCD calculations for the $J/\psi$ - not at SPS nor
at RHIC. Then the observed suppression comes from screening of the higher-mass states
alone ($\psi\prime$ and $\chi_C$) that, by their decay, normally provide $\sim$40\% of the observed $J/\psi$s. This
scenario is consistent with the apparently identical suppression patterns seen at
the SPS and RHIC shown in Fig.~\ref{fig:satz_RHIC_SPS_energy_density}.

As a result we are left for the moment with two different scenarios that provide
explanations for the RHIC A+A data. Both include the QGP in their picture, either through
color screening in the QGP or through severe suppression of the $J/\psi$ by a very high gluon
density. Further tests from the data will be necessary to clarify the picture.
Regeneration models predict narrowing of both the rapidity and $p_T$ distributions, but so
far the preliminary data shows little or no change in the rapidity shape from ordinary
p+p and only a hint of narrowing of the $p_T$. We are also trying to extract a measurement
of flow for the $J/\psi$, since emerging results for single charm are beginning to show flow 
and the $J/\psi$'s, if they were from regeneration, would inherit this flow. These tests await
the more precise final analysis of the 2004,5 Au+Au and Cu+Cu data; and higher statistics
runs for Au+Au and d+Au in the near future.

\begin{ltxfigure}[tbh]
\begin{minipage}[b]{0.46\linewidth}
  \centering
  \includegraphics[width=0.9\textwidth]{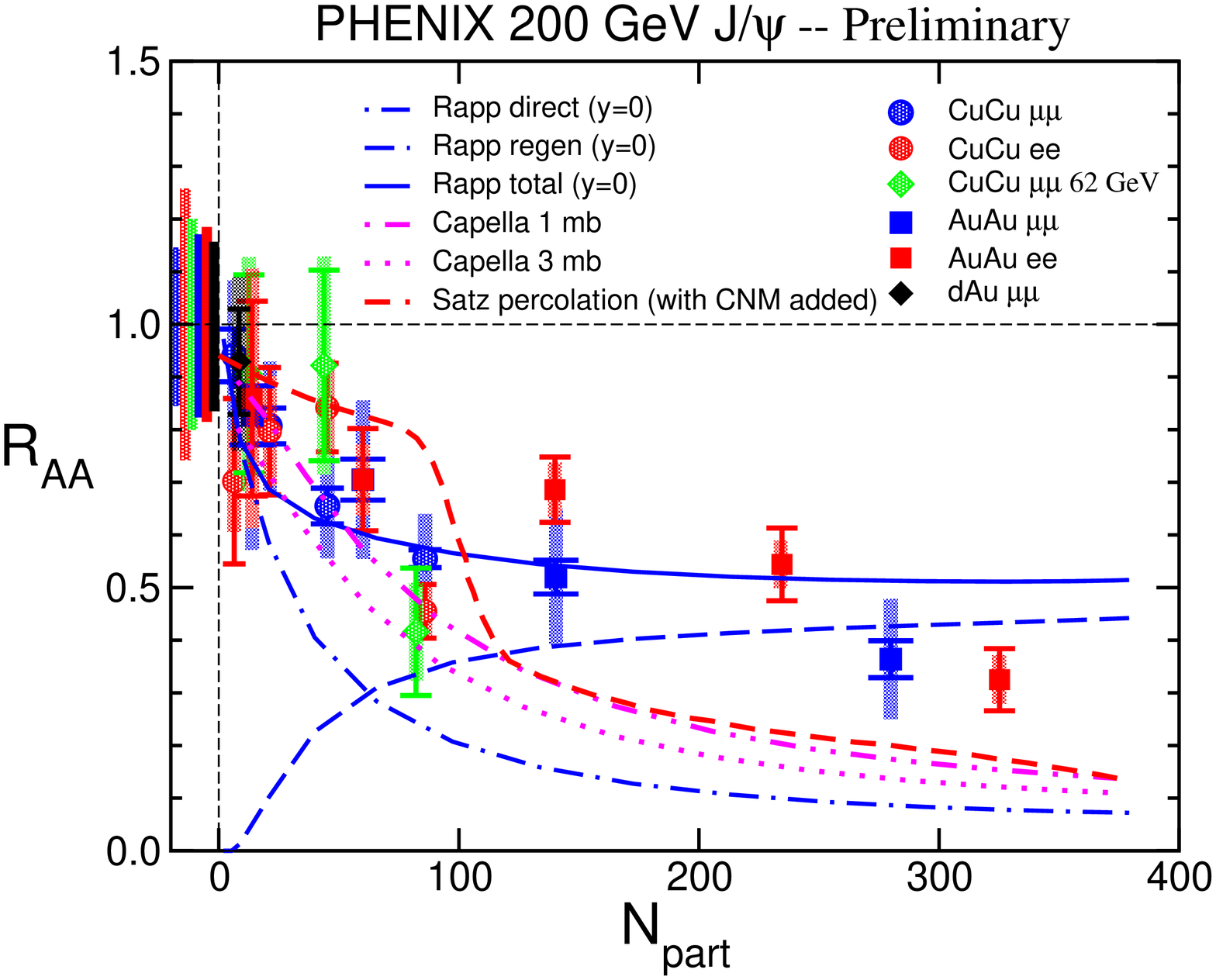}
  \caption{Theories that agree with SPS data do not agree with RHIC data,
  unless regeneration is added as in the Rapp (solid blue) curve.}
  \label{fig:raa_npart_supp_rapp}
\end{minipage}
\hspace{0.2cm}
\begin{minipage}[b]{0.46\linewidth}
  \includegraphics[width=0.75\textwidth]{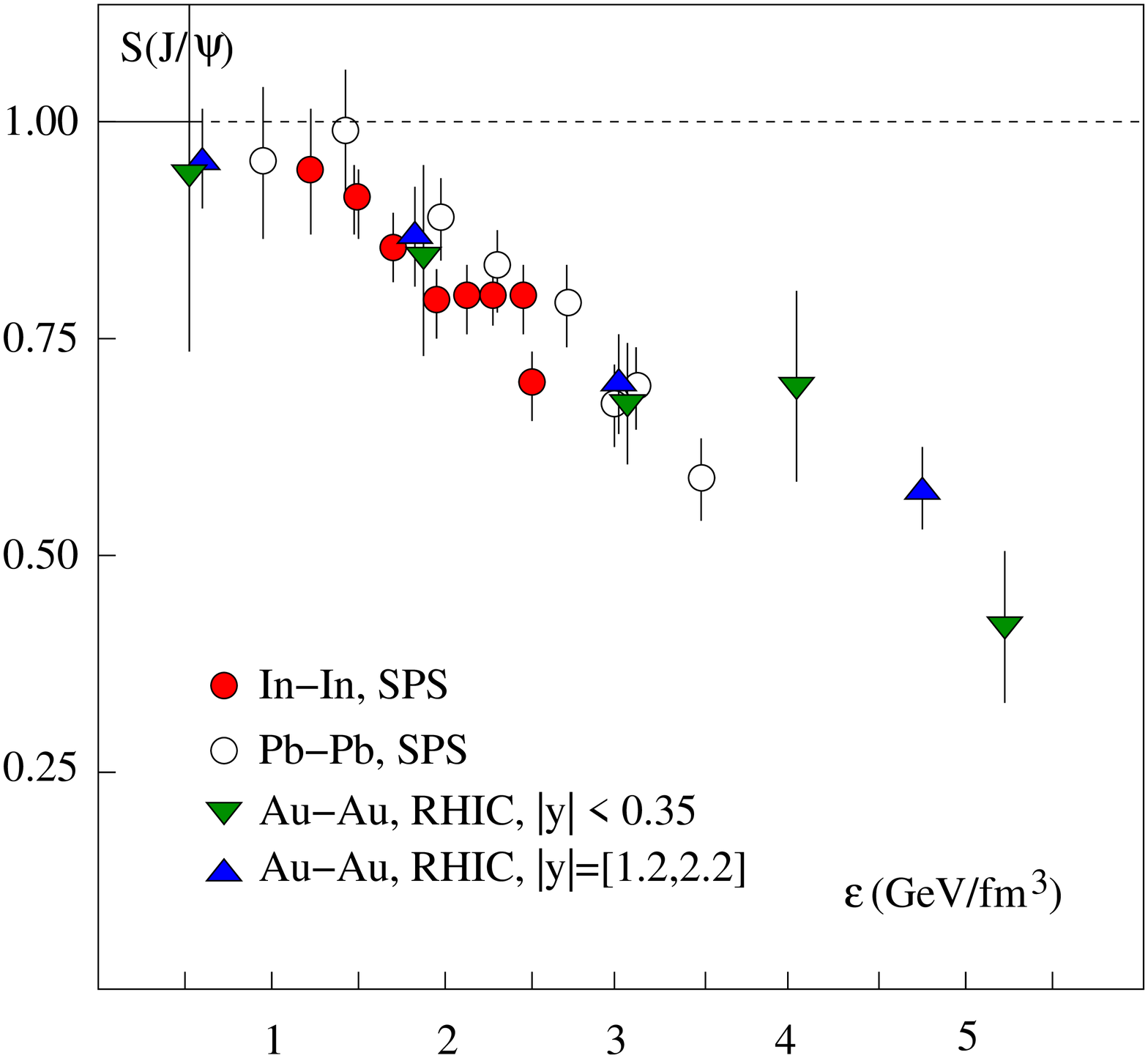}
  \caption{Universal dependence on energy density of $J/\psi$ suppression measurements
  at RHIC and at the SPS\cite{karsch}.}
  \label{fig:satz_RHIC_SPS_energy_density}
\end{minipage}
\end{ltxfigure}

\section{Open Charm in Au+Au Collisions}

In Au+Au collisions, open charm (and beauty) together, are suppressed due to energy loss
in the dense medium, with gluon densities per unit rapidity of up to 1000 infered in some
theoretical analysis. However, as shown in Fig.~\ref{fig:charm_dedx_wicks}, calculations
that include both radiative and collisional energy loss\cite{wicks} predict too small a suppression
when both charm and beauty are included. Flow has also been oberved for heavy quark production.
As shown in  Fig.~\ref{fig:charm_flow}, the flow is similar to that of light quarks at small $p_T$,
but at higher $p_T$ the data with large uncertainties hints at vanishing flow, consistent with
simple expectations that higher $p_T$ charm simply punches out of the medium and never thermalizes.

\section{Summary}

Substantial uncertainties remain in the understanding of charm production cross sections, and the
polarization of charmonia. There are also a number of cold nuclear matter effects that influence
their production in nuclei and cloud our understanding of the suppression seen in nucleus-nucleus
collisions. Two competing pictures are able to explain the $J/\psi$ suppression seen in
nucleus-nucleus collisions at RHIC - one involving sequential screening in the plasma of the various
charmonia states; the other with strong dissociation of all charmonia states by a dense gluon field
but recombination of independently produced charm quarks. For
open charm, the the energy loss observed in the dense medium from nucleus-nucleus collisions is
larger than that expected from theoretical models that include radiative and collisional energy loss
of both charm and beauty. Higher statistics data with higher luminosity runs as well as RHIC vertex
detector upgrades will enable more precise data in the future that will give a clearer understanding
of the rich physics in charm production.

\begin{ltxfigure}[tbh]
\begin{minipage}[b]{0.42\linewidth}
  \centering
  \includegraphics[width=0.8\textwidth]{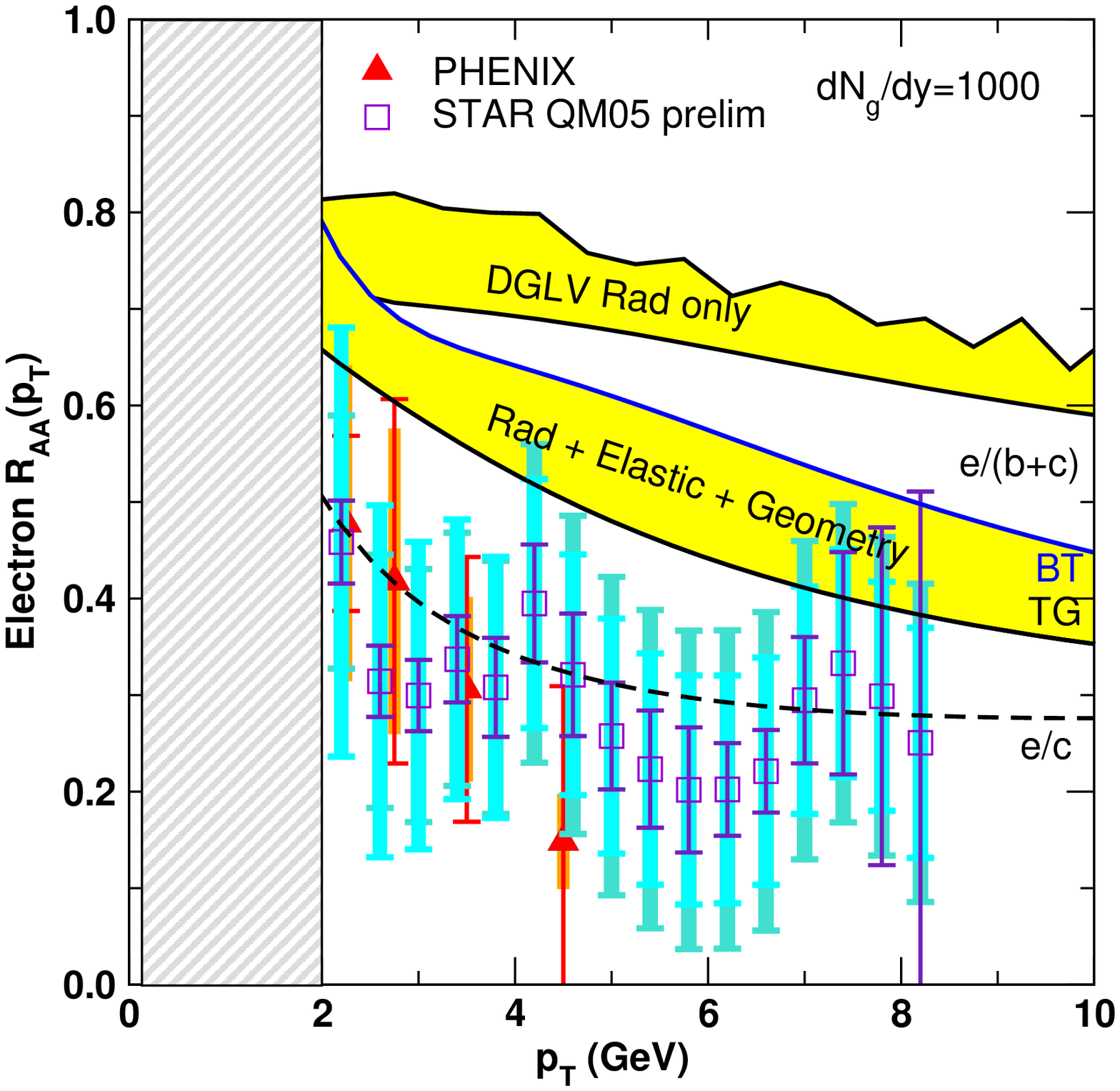}
  \caption{Energy loss calculations compared to open heavy (charm + beauty) data vs $p_T$\cite{wicks}.}
  \label{fig:charm_dedx_wicks}
\end{minipage}
\hspace{0.2cm}
\begin{minipage}[b]{0.42\linewidth}
  \includegraphics[width=\textwidth]{figures/charm_flow.eps}
  \caption{Elliptic flow of open heavy (charm + beauty) compared to Rapp calculations\cite{rapp_flow}.}
  \label{fig:charm_flow}
\end{minipage}
\end{ltxfigure}




\bibliographystyle{aipproc}   





\end{document}